\documentclass[preprint,12pt]{elsarticle}

\usepackage{bm}
\usepackage{amsmath,amssymb,amsfonts,times}
\usepackage{url}

\usepackage[colorlinks=true, pdfstartview=FitV, linkcolor=red, citecolor=blue, urlcolor=blue]{hyperref}

\usepackage{graphicx}
\usepackage{epsfig}
\usepackage{slashed}
\usepackage{a4wide}
\usepackage{fancyhdr}
\usepackage{subfigure}

\usepackage{rotating}

\usepackage{pdflscape}

\allowdisplaybreaks[1]

\begin{document}

\begin{frontmatter}

\title{Machine learning light hypernuclei}

\author{Isaac Vida\~na}
\address{Istituto Nazionale di Fisica Nucleare, Sezione di Catania, Dipartimento di Fisica ``Ettore Majorana'', Universit\`a di Catania, Via Santa Sofia 64, I-95123 Catania, Italy}


\begin{abstract}
We employ a feed-forward artificial neural network to extrapolate at large model spaces the results of {\it ab-initio} hypernuclear No-Core Shell Model calculations for the $\Lambda$ separation energy $B_\Lambda$ of the lightest hypernuclei, $^3_\Lambda$H, $^4_\Lambda$H and $^4_\Lambda$He, obtained in computationally accessible harmonic oscillator basis spaces using chiral nucleon-nucleon, nucleon-nucleon-nucleon and hyperon-nucleon interactions. The overfitting problem is avoided by enlarging the size of the input dataset and by introducing a Gaussian noise during the training process of the neural network. We find that a network with a single hidden layer of eight neurons is sufficient to extrapolate correctly the value of the $\Lambda$ separation energy to model spaces of size $N_{max}=100$. The results obtained are in agreement with the experimental data in the case of $^3_\Lambda$H and the $0^+$ state of $^4_\Lambda$He, although they are off of the experiment by about $0.3$ MeV for both the $0^+$ and $1^+$states of $^4_\Lambda$H and the $1^+$ state of $^4_\Lambda$He. We find that our results are in excellent agreement with those obtained using other extrapolation schemes of the No-Core Shell Model calculations, showing this that an ANN is a reliable method to extrapolate the results of hypernuclear No-Core Shell Model calculations to large model spaces.
\end{abstract}


\end{frontmatter}

\section{Introduction}
Artificial intelligence (AI) and, in particular, machine learning (ML) have become one of the most exciting 
and dynamic areas of research in recent years, impacting many domains of science and technology \cite{Bishop06,Hastie09,Murphy12,LeCun15,Goodfellow16}. The main challenge in ML is to devise 
algorithms able to recognize patterns in previously unseen data without any explicit instructions by an external party. Among the different existing ML algorithms, artificial neural networks (ANNs) \cite{McCulloch43} have currently turned into the most largely ones used in science. Inspired by biological neural networks, the architecture of ANNs consists of an input layer, one or more hidden layers, and an output layer of several interconnected artificial neurons. Combining this architecture with various training algorithms such as back-propagation and stochastic gradient descent, ANNs are able to capture complicated non-linear input-output relationships in datasets. Due to this ability, ANNs can be considered as universal non-linear function approximators \cite{Hornik89,Park91,Scarselli98}. Examples of ANNs include among others feed-forward neural networks, convolutional neural networks, recurrent neural networks, generative adversarial networks, radial basis function networks, Boltzmann machines or restricted Boltzmann machines (see {\it e.g.}, Ref.\ \cite{Goodfellow16} for a detailed description of these and other types of ANNs).


In physics ML and ANNs have been applied in condense matter, statistical physics, cold atoms, quantum many-body theory, quantum computing, cosmology, particle physics or nuclear physics (see Refs.\ \cite{Mehta19,Carleo19,Deiana21} and references therein for recent reviews on the application of ML and ANNs in physics). In particular, applications in nuclear physics \cite{Bedaque21,Boehnlein21} date back to the beginning of the 90's when Gazula {\it et al.,} \cite{Gazula92} employed a 
feed-forward neural network to study global nuclear properties across the nuclear landscape. Since then, ML and ANNs have been used to predict among other things nuclear masses \cite{Athanassopoulos04,Utama16,Niu18,Carnini20,Wu20,Wu21,Yuksel21,Liu21,Gao21}, charge radii \cite{Akkoyun13,Utama16b,Ma20,Wu20b,Dong22}, $\alpha$- and $\beta$-decay half-lives \cite{Freitas19,Saxena21,Costiris09,Niu19,Rodriguez19}, fission yields \cite{Wang19,Lowell19,Lowell20,Qiao21,Wang21}, fusion reaction cross-sections \cite{Akkoyun20}, isotropic cross-sections in proton-induced spallation reactions \cite{Ma20b,Ma20c}, ground-state and excited energies \cite{Lasseri20}, dripline locations \cite{Neufcourt18,Neufcourt19}, the deuteron properties \cite{Keeble20},
proton radius \cite{Graczyk14,Graczyk15}, the nuclear liquid-gas phase transition \cite{Wang20}, nuclear energy density functionals \cite{Wu21b}, neutron star properties and the nuclear matter equation of state \cite{Fujimoto18,Fujimoto20,Morawski20,Anil20,Ferreira21}, or to extract the nucleon axial form factor from the analysis of neutrino-scattering data \cite{Alvarez-Ruso19}. 

Recently, ANNs have been also employed to extrapolate the results of {\it ab initio} nuclear structure calculations in finite model spaces, which otherwise would be extremely costly from a computational point of view. In particular, Negoita {\it et al.} \cite{Negoita18,Negoita19} have used a feed-forward ANN method for predicting the ground state energy and the ground state point proton root-mean-squared radius of $^6$Li. Training the network with No-Core Shell Model (NCSM) results, obtained in accessible harmonic oscillator (HO) basis spaces for various oscillators spacings $\hbar\omega$ and different numbers $N_{max}\hbar\omega$ of maximum excitation energies, these authors showed that an ANN is able to predict correctly extrapolations of the NCSM results to very large model spaces of size $N_{max}\sim 100$, and that their dependence on $\hbar\omega$ vanishes as the value of $N_{max}$ increases. Similarly, Jiang {\it et al.} \cite{Jiang19} have also employed an ANN to extrapolate the ground state energy and radii of $^4$He, $^6$Li and $^{16}$O computed with the NCSM and the coupled-cluster (CC) methods. These authors have shown that a preprocessing of the input data, and the inclusion of correlations among these data, reduces the problem of multiple solutions yielding more stable results and consistent estimations of the uncertainty. 

Following the works of Negoita {\it et al.} \cite{Negoita18,Negoita19} and Jiang {\it et al.} \cite{Jiang19}, in this paper we employ a feed-forward ANN to extrapolate, at large model spaces, the hypernuclear NCSM results of Refs.\ \cite{Wirth18,Htun21,Gazda22} for the 
$\Lambda$ separation energies $B_\Lambda$ (defined as the difference between the binding energies of the hypernucleus and the corresponding core nucleus) of the lightest hypernuclei, $^3_\Lambda$H, $^4_\Lambda$H and $^4_\Lambda$He, obtained with chiral nucleon-nucleon \cite{Entem03} and nucleon-nucleon-nucleon \cite{Navratil07} interactions at N$^3$LO and N$^2$LO, respectively, both with a regulator cutoff of $500$ MeV, and hyperon-nucleon \cite{Polinder06} potentials at LO with a cutoff of $600$ MeV. We find that an ANN with a single hidden layer of eight neurons is sufficient to extrapolate correctly the  
$\Lambda$ separation energies of the three hypernuclei considered. This is in agreement with the {\it universal approximation theorem} \cite{Cybenko89,Funahashi89,Hornik91} which assures that any continuous function can be realized by a neural network with just one hidden layer.

The manuscript is organized in the following way. A brief description of feed-forward ANNs and, particularly, of the one employed in this work is presented in Sec.\ \ref{sec:architecture}. The results of the extrapolation to large model spaces of the $\Lambda$ separation energy of $^3_\Lambda$H, $^4_\Lambda$H and $^4_\Lambda$He are shown and discussed in
Sec.\ \ref{sec:results}. Finally, a short summary and the conclusions of this work are given in Sec.\ \ref{sec:summ_conclu}. 
 
\section{Feed-forward Artificial Neural Networks}
\label{sec:architecture}

ANNs consists of a series of layers (input, hidden and output) each one containing a certain number of interconnected nodes called neurons. In particular, a feed-forward ANN is a type of ANN wherein connections between the neurons do not form a cycle, and the data propagates sequentially from the input to the output layer through all the hidden layers. At each one of the $N_k$ neurons $i$ of a given layer $k$, the set of input data $\{x_j^{(k-1)}\}$ from the $N_{k-1}$ neurons $j$ of the previous layer $k-1$ is transformed into 
\begin{equation}
x_i^{(k)}=\sigma^{(k)}\left(\sum_{j=1}^{N_{k-1}}W_{ij}^{(k)}x_j^{(k-1)}+a_i^{(k)}\right) \ .
\label{eq:transform}
\end{equation}
Note that for the input layer (which is labelled 0 in this work) one has simply $x_i^{(0)}=x_i$ ($1\leq i \leq N_0$), where $x_i$ is the input dataset. In Eq.\ (\ref{eq:transform}), $\sigma^{(k)}(z)$ is the so-called {\it activation function} responsible for the introduction of non-linearities on the neural network that enable it to capture complex non-linear relationships in the dataset. There exist several possible choices for the activation function depending on the particular problem (classification or regresion) one is trying to solve. The choice of the activation function has a large impact on the capability and performance of the neural network, and different activation functions may be used in different parts of the model. Some common choices include, among other activation functions, the sigmoid one $\sigma(z)=1/(e^x+1)$, the Rectified Linear Unit (ReLu) $\sigma(z)=\,$max$\{0,z\}$ or the hyperbolic tangent $\sigma(z)=\,$tanh$(z)$. Particularly, in this work we are solving a regression-type problem and we employ the sigmoid function in the hidden and output layers. The coefficients $W_{ij}^{(k)}$ and $a_i^{(k)}$ in Eq.\ (\ref{eq:transform}) are the fitting parameters of the network and denote, respectively, the {\it weights} of the connections between the neurons of the two adjacent layers $k-1$ and $k$, and the activation offset, called {\it bias}, of each neuron of the layer $k$. The total number of fitting parameters $n_{p}$ is given by 
\begin{equation}
n_{p}=\sum_{k=0}^{L-2}(N_k+1)N_{k+1} \ ,
\label{eq:totpar}
\end{equation}
where $L$ is the total number of layers of the network (input, hidden and output), and $N_k$ and $N_{k+1}$ are the number of neurons in the layers $k$ and $k+1$, respectively. The learning process of an ANN involves the minimization of a loss function in order to obtain the optimal set of fitting parameters 
$({\bf W,a})\equiv\{W^{(k)}_{ij}, a^{(k)}_i\}$. As in the case of the activation function, the choice of the loss function depends on the type of problem one is solving with a neural network. In the present work for the loss function we chose the mean squared error (MSE), a common choice in the case of regression-type problems
\begin{equation}
{\cal L}({\bf W,a})=\frac{1}{N}\sum_{i=1}^N\left({\hat y}_i({\bf W,a})-y_i\right)^2 \ ,
\label{eq:mse}
\end{equation}
with $N$ being the number of data points used in the minimization procedure, ${\hat y_i}({\bf W,a})\equiv x_i^{(L)}$ (see Eq.\ (\ref{eq:transform})) the prediction of the network, and $y_i$ the actual output of the input data, in our case the $\Lambda$ separation energy.
 
\begin{center}
\begin{figure}[t]
\begin{center}
\includegraphics[width=0.45\textwidth,keepaspectratio]{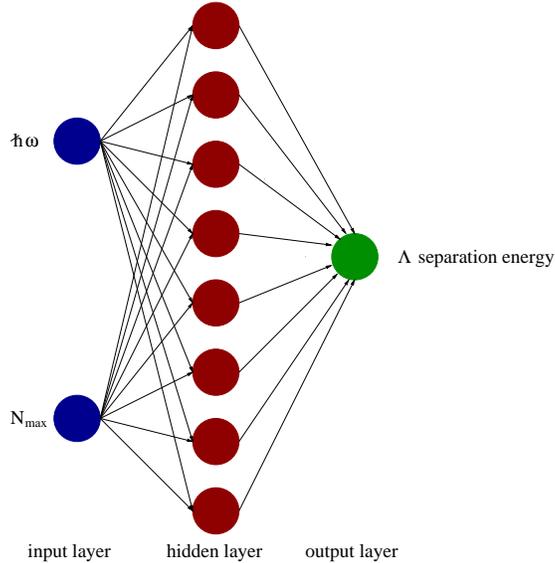}
\caption{(Color online). Architecture of the feed-forward ANN employed in this work.
The input data are the HO spacings $\hbar\omega$ and the maximum number of basis states 
$N_{\max}$ employed in the hypernuclear NCSM calculations of Refs.\ \cite{Wirth18,Htun21,Gazda22}, whereas the output is the $\Lambda$ separation energy $B_\Lambda$ of the corresponding light hypernucleus under study.}
\label{fig:fig1}
\end{center}
\end{figure}
\end{center}

A major issue in the development of an ANN is overfitting (also known as overtraining), which basically means that the network, due to its high flexibility to approximate complex non-linear functions, tries to fit the training data entirely, and ends up memorizing all the data patterns. As a consequence the predictability of the neural network on testing data in this case becomes questionable. Several strategies can be followed to avoid this problem such as {\it e.g.,} the early stopping of the training \cite{Prechelt98} or the use of regularization techniques \cite{Zou05} like the dropout 
one \cite{Srivastava14}. In addition to these strategies, which can be used together to increase the network robustness, overfitting can be reduced by enlarging the dataset \cite{Tanner87}. Another important question is how do neural networks extrapolate, {\it i.e.,} how do they perform beyond the range of the training dataset. Earlier works showed that feed-forward ANNs fail to extrapolate well in certain simple tasks \cite{Barnard92,Haley92}. However, it  was shown later that graph neural networks, a certain class of structured networks with feed-forward ANNs as building blocks, extrapolate with some success in more complex tasks \cite{Battaglia16,Velickovic20,Lample20}. Very recently, the authors of Ref.\ \cite{Xu21} have identified the conditions under which feed-forward and graph neural networks extrapolate as desired. The analysis of these conditions is, however, out of the scope of the present work, and the interested reader is referred to the work of these authors for details.

The architecture of the neural network employed in this work is shown in Fig.\ \ref{fig:fig1}. It consist of 3 layers with 2, 8 and 1 neurons in the input, hidden and output layers, respectively. The choice of this simple architecture is guided by the {\it Ockham's razor} or {\it parsimony principle} in the sense that if two models have the same performance
on the validation/testing dataset, the simpler model is preferred over the more complex one, and by the {\it universal approximation theorem} \cite{Cybenko89,Funahashi89,Hornik91} which, as said already in the introduction, states that a neural network with one hidden layer can approximate any continuous function for inputs within a specific range. We have checked that, in fact, a larger number of hidden layers decreases the performance of the network. Then, according to Eq.\ (\ref{eq:totpar}), there are 33 fitting parameters (24 weights and 9 biases) in our neural network. To optimize the fitting parameters of our network we employ the Adam algorithm \cite{Kingma14}, an extended version of the stochastic gradient descent method \cite{Ruder16}. The input dataset of our network consist of two features, the HO spacings $\hbar\omega$ and the maximum number of HO basis states $N_{\max}$ employed in the hypernuclear NCSM calculations of Refs.\ \cite{Wirth18,Htun21,Gazda22}, and the corresponding target observable, namely, the $\Lambda$ separation energy of the light hypernuclei under study. We use the $80\%$ of the input dataset to train the network and leave the $20\%$ of it to test it. Since the available input dataset in our case is not too large (we have only 128 data points in the case of $^3_\Lambda$H and much less in the $^4_\Lambda$H and $^4_\Lambda$He cases, see Refs.\ \cite{Wirth18,Htun21,Gazda22}), to avoid overfitting we enlarge it by performing a cubic spline interpolation in the HO spacing $\hbar\omega$ at each given value of $N_{max}$ of the original input dataset of Refs.\ \cite{Wirth18,Htun21,Gazda22}. To further reduce the overfitting, we also introduce a Gaussian noise in the enlarged input dataset during the training of the network. Adding noise makes the network less able to memorize data patterns since they change randomly during all the training process and, consequently, overfitting is largerly reduced. The enlargement of the input dataset and the addition of noise are simple and economical ways to obtain more information from the limited amount of existing input data, making our basic single-hidden-layer network more robust. In addition, we have employed a $10\%$ of the training data subset to give an estimate of the network skill while tuning its hyperparameters ({\it i.e.}, number of layers and neurons, activation function, optimizer algorithm, number of iterations during the training process, ...) to arrive to our final model. Note that this validation dataset is different from the test dataset which is used to give an unbiased estimate of the skill of the final tuned network.

As a final remark of this section, we would like to mention that the numerical implementation of our feed-forward ANN have been done with the Python \cite{Python} libraries Scikit-learn \cite{Scikit} and Keras \cite{Keras} using a TensorFlow \cite{TensorFlow} backend.

\begin{center}
\begin{figure}[t]
\begin{center}
\includegraphics[width=0.45\textwidth,keepaspectratio]{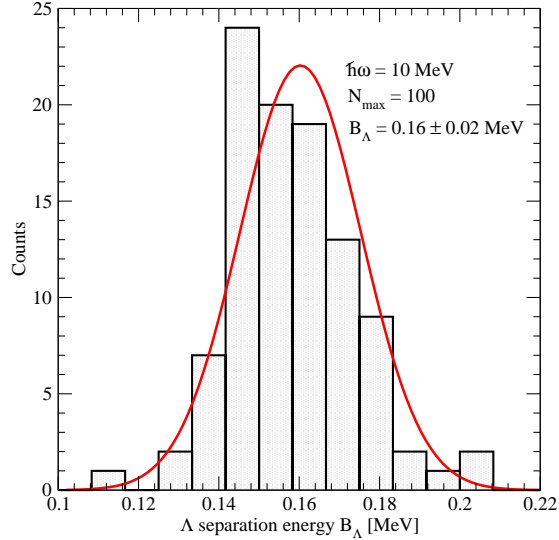}
\caption{(Color online). Statistical distribution of the results for the
$\Lambda$ separation energy $B_\Lambda$ of the ground state of $^3_\Lambda$H predicted by 100 independent
runs of the ANN for an HO spacing $\hbar\omega=10$ MeV and a model space size $N_{max}=100$. The continuous line shows the Gaussian fit of the histogram. The average value of $B_\Lambda$ and it corresponding error (standard deviation) is indicated.}
\label{fig:fig2}
\end{center}
\end{figure}
\end{center}

\begin{center}
\begin{figure*}[t]
\begin{center}
\includegraphics[width=0.45\textwidth,keepaspectratio]{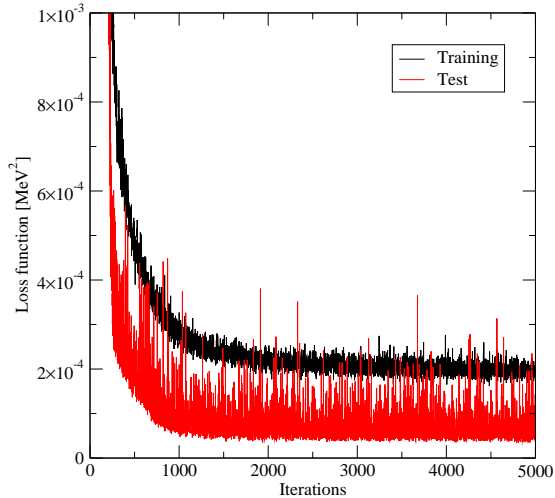}
\caption{(Color online). Loss function ${\cal L}({\bf W,a})$ of the training and test datasets as a function of the number of iterations in the calculation of the $\Lambda$ separation energy of the ground state of $^3_\Lambda$H.}
\label{fig:fig3}
\end{center}
\end{figure*}
\end{center}

\section{$\Lambda$ separation energy in light hypernuclei}
\label{sec:results}

In this section we finally present the results of the extrapolation to large model spaces of the $\Lambda$ separation energy $B_\Lambda$ in $^3_\Lambda$H, $^4_\Lambda$H and $^4_\Lambda$He, predicted by our ANN. We should note that in general a typical run of an ANN starts with random values of the weights and biases of the network. The random initialization of the weights and biases is not accidental but an important feature of the network training that introduces in it a certain degree of stochasticity (in addition to the one introduced by the use of optimization algorithms such as the stochastic gradient descent or one of its extensions) which reduces the risk that during the optimization process of the network parameters it gets stuck in a local minimum. Consequently, different runs of the ANN lead to slightly different results, as it can be seen for instance in Fig.\ \ref{fig:fig2} where
it is shown the statistical distribution of the results for the $\Lambda$ separation energy $B_\Lambda$ of the ground state of $^3_\Lambda$H predicted by 100 independent runs of the ANN for an HO spacing $\hbar\omega=10$ MeV and a model space of size $N_{max}=100$. We note that, for each hypernucleus and each state considered in this work, we have taken the average value and the standard deviation of 100 independent runs of the ANN as the predictions of the network and their corresponding error, respectively.  

\begin{center}
\begin{figure*}[t]
\begin{center}
\includegraphics[width=0.8\textwidth,keepaspectratio]{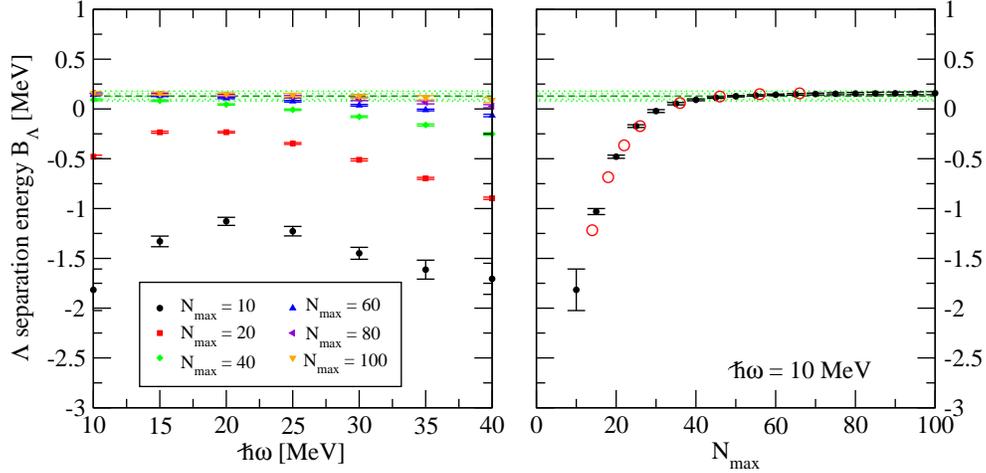}
\caption{(Color online). $\Lambda$ separation energy $B_\Lambda$ of the ground state of $^3_\Lambda$H as a function of the HO spacing $\hbar\omega$ (left panel) for several values of the model space size $N_{max}$, and as a function of $N_{max}$ (right panel) for the HO spacing $\hbar\omega=10$ MeV. The experimental value with its corresponding error taken from Refs.\ \cite{Davis05,Gal16} are shown by the dashed line and the colored band, respectively. The open circles in the right panel show the NCSM results of Refs.\ \cite{Htun21,Gazda22} used for the training of the neural network.}
\label{fig:fig4}
\end{center}
\end{figure*}
\end{center}

\begin{center}
\begin{figure*}[t]
\begin{center}
\includegraphics[width=0.8\textwidth,keepaspectratio]{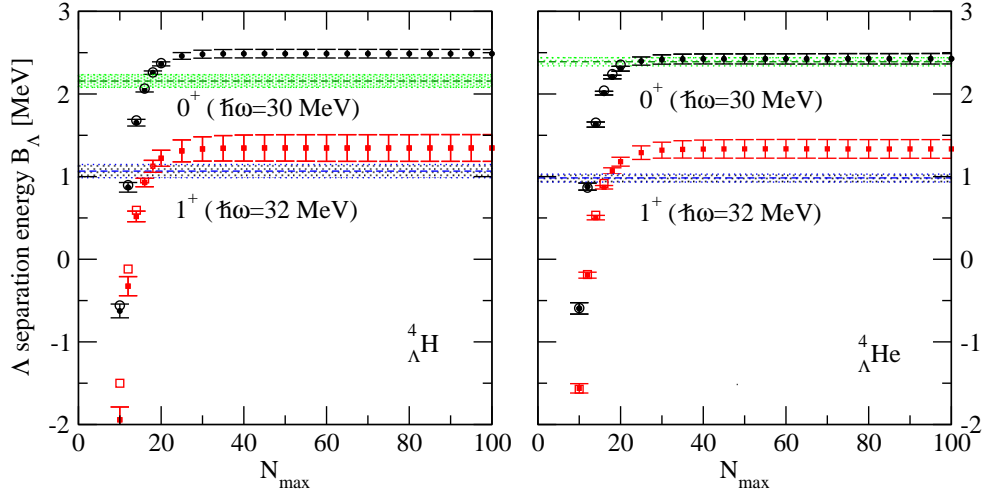}
\caption{(Color online). $\Lambda$ separation energy $B_\Lambda$ of the $0^+$ and $1^+$ states of $^4_\Lambda$H (left panel) and $^4_\Lambda$He (right panel) as a function of the model space size $N_{max}$ for the HO spacing $\hbar\omega=30(32)$ MeV, considered in Refs.\ \cite{Wirth18,Gazda22} the optimal one for the $0^+(1^+)$ state of $^4_\Lambda$H and $^4_\Lambda$He. The experimental values with their corresponding errors taken from Refs.\ \cite{Davis05,Schulz16,Yamamoto15} are shown by the dashed lines and the colored bands, respectively. The open symbols in both panels show the NCSM results of Refs.\ \cite{Wirth18,Gazda22} used for the training of the neural network.}
\label{fig:fig5}
\end{center}
\end{figure*}
\end{center}

\begin{center}
\begin{figure}[t]
\begin{center}
\includegraphics[width=0.7\textwidth,keepaspectratio]{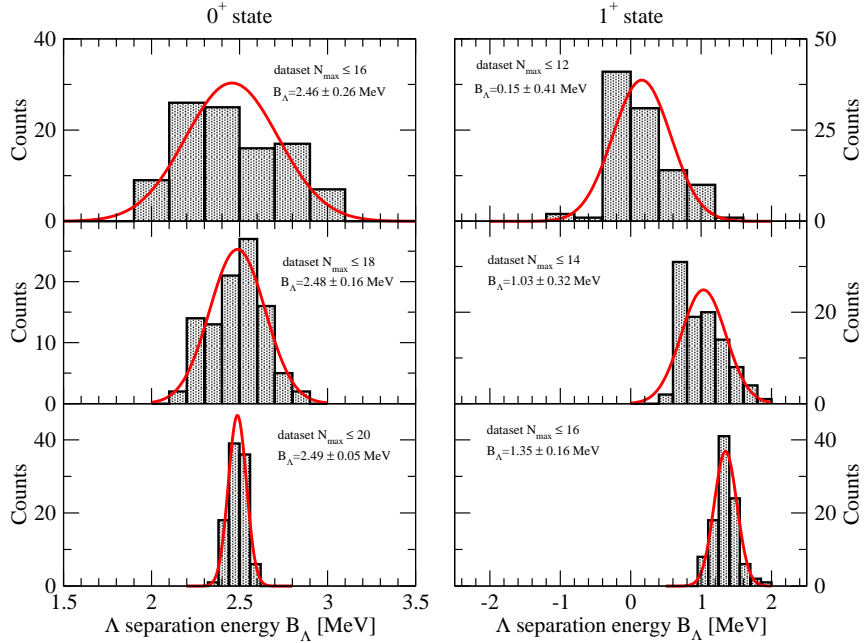}
\caption{(Color online). Statistical distribution of the results for the
$\Lambda$ separation energy $B_\Lambda$ of the $0^+$ (left panels) and $1^+$ (right panels) states of $^4_\Lambda$H predicted by 100 independent runs of the ANN for several choices of the maximal value of $N_{max}$
included in the training dataset. Results are shown for an HO spacing $\hbar\omega=30(32)$ MeV for the $0^+$($1^+$) state and a model space size $N_{max}=100$. The continuous lines show the Gaussian fits of the histograms. The average value of $B_{\Lambda}$ and its corresponding error (standard deviation) is indicated in each panel.}
\label{fig:fig6}
\end{center}
\end{figure}
\end{center}

\begin{center}
\begin{figure}[t]
\begin{center}
\includegraphics[width=0.7\textwidth,keepaspectratio]{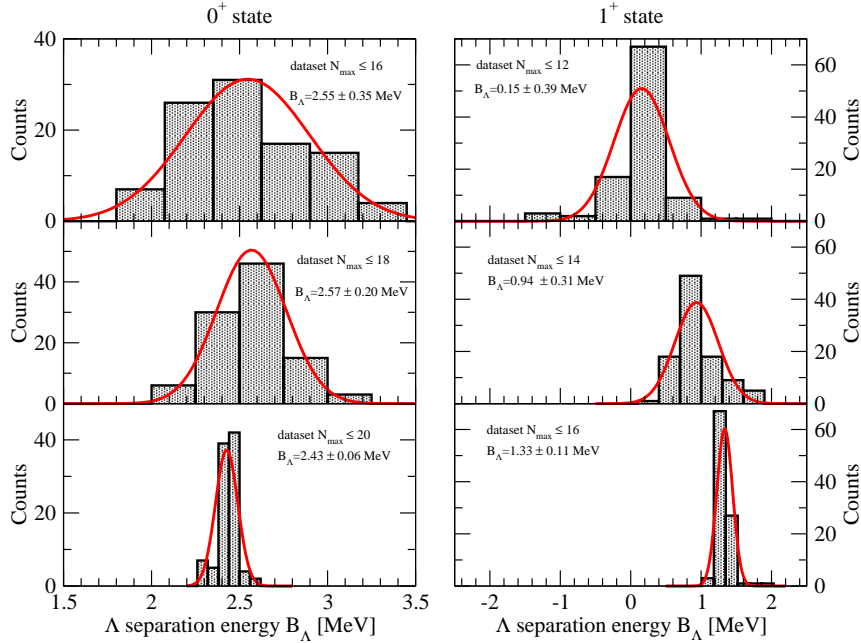}
\caption{(Color online). Same as Fig.\ \ref{fig:fig6} for the $0^+$ and $1^+$ states of $^4_\Lambda$He.}
\label{fig:fig7}
\end{center}
\end{figure}
\end{center}

Before discussing the results, to illustrate the network performance, we show in Fig.\ \ref{fig:fig3} the loss function ${\cal L}({\bf W,a})$ of the training and test datasets as a function of the number of iterations in the calculation of the $\Lambda$ separation energy of the ground state of $^3_\Lambda$H. As mentioned before, the training (test) dataset is made of the $80\%$ ($20\%$) of the enlarged input dataset which includes, as we also said, a Gaussian noise. As it is seen the loss function of both training and test datasets decreases very fast during the first 500 iterations and it becomes (on average) essentially quite constant at about 1000 iterations and above it. This illustrates the good performance of the Adam optimization algorithm used in the present work. In addition, the loss function of the test dataset is smaller than that of the training one, indicating that overfitting has been avoided. A similar good performance of our ANN is obtained also for the predictions of the $\Lambda$ separation energies of $^4_\Lambda$H and $^4_\Lambda$He.

The prediction of our ANN for the $\Lambda$ separation energy of the ground state of $^3_\Lambda$H is shown in the left panel of 
Fig.\ \ref{fig:fig4} as a function of the HO spacing $\hbar\omega$ for several values of the model space size $N_{max}$ and, in the right panel, as a function of $N_{max}$ for $\hbar\omega=10$ MeV. As we said at the beginning of this section, we consider as the prediction of our ANN the average value of the results obtained in 100 independent runs of the network, and the standard deviation of these results
as its associated error, which is also shown in the figure (note that the size of the error bars is in general smaller than that of the symbol). The experimental value and its error ($B_\Lambda^{exp}$$(^3_\Lambda$H$)=0.13\pm0.05$ MeV \cite{Davis05,Gal16}) are shown in both panels of the figure by the dashed line and the colored band, respectively. The open circles in the right panel of the figure show the NCSM results of Refs.\ \cite{Htun21,Gazda22} used during the training process of the ANN. We note that the values of the HO spacing and the model space size included in the input dataset are in the ranges $\hbar\omega=10-40$ MeV and $N_{max}=14-70$, respectively. The left panel shows the convergence of $B_\Lambda$ with both basis space parameters, $\hbar\omega$ and $N_{max}$. Note that this convergence is very slow mainly due to the extremely weak binding energy of the $^3_\Lambda$H ($E_{exp}$$(^3_\Lambda$H$)=-2.35\pm 0.01$ MeV). As it is also seen, the dependence of  $B_\Lambda$ with $\hbar\omega$ reduces considerably with the increase of $N_{max}$ and, although, some little dependence still remains for $N_{max}=100$, it is expected to vanish completely for larger values of $N_{max}$. This was already pointed out by Negoita {\it et al.} in their ANN study of $^6$Li \cite{Negoita18,Negoita19}. We note that this is neither a coincidence nor a particular feature of ANNs but it is actually the expected behaviour of NCSM calculations (see for instance Ref.\ \cite{Barrett13}) and, therefore, the observed reduction of
the dependence on the HO spacing denotes that the ANN catches properly the behaviour of the input data. In the right panel it can be seen that our ANN does not only give a good prediction of $B_\Lambda$ in the region of values of $N_{max}$ used in the NCSM calculations of Refs.\ \cite{Htun21,Gazda22}, but it also extrapolates reasonably well to large values of the model space size. In particular, for $N_{max}=100$ our ANN predicts $B_\Lambda(^3_\Lambda$H$)=0.16 \pm 0.02$ MeV, a value compatible with the experimental error.

The $\Lambda$ separation energies of the $0^+$ and $1^+$ states of $^4_\Lambda$H and $^4_\Lambda$He are shown, respectively, in the left and right panels of Fig.\ \ref{fig:fig5} as a function of $N_{max}$ for $\hbar\omega=30(32)$ MeV, considered in Refs.\ \cite{Wirth18,Gazda22} the optimal value of the HO spacing for the $0^+(1^+)$ state of $^4_\Lambda$H and $^4_\Lambda$He.  As in the case of
$^3_\Lambda$H, the experimental values with their corresponding errors are shown by the dashed lines and the colored bands, respectively, whereas the NCSM results of Refs.\ \cite{Wirth18,Gazda22}, used in the training of the network, are shown 
by the open symbols. We note that the values of $\hbar\omega$ included in the
input dataset are in the range $28-34$ MeV for both states of these two hypernuclei, whereas those of $N_{max}$ are within the values $2-20$ and $2-16$ for the $0^+$ and 
$1^+$ state, respectively. We note that also in this case we have performed, for each one of these two hypernuclei and each one of their states, 100 independent runs of the network and we have taken the average value and the standard deviation of the results of these runs as the predictions of our ANN for the $\Lambda$ separation energies and their errors. Notice that for these two hypernuclei, which are more heavier and bound systems, the convergence of the results with the size of the model space is much faster than in the case of $^3_\Lambda$H and, in fact, a quite good converged result is obtained already for $N_{max}\gtrsim 25$. To further check the convergence of the extrapolated value of $B_\Lambda$ for the $0^+$ and $1^+$ states of both hypernuclei at $N_{max}=100$, we show in Figs.\ \ref{fig:fig6} and \ref{fig:fig7} the statistical distributions of the results of 100 independent runs of the ANN for several choices of the maximal value of $N_{max}$ taken into account in the training dataset. In both figures the results are shown for an HO spacing $\hbar\omega=30(32)$ MeV for the $0^+$($1^+$) state. The continuous lines show in each case the corresponding Gaussian fits of the histograms, and the average value of $B_{\Lambda}$ together with its corresponding error (standard deviation) are indicated in each panel of both figures. Note that when the maximal value of $N_{max}$ included in the training dataset is increased, the dispersion of the results predicted by the ANN reduces and their corresponding distributions becomes narrower and narrower. The increase of the maximal value of $N_{max}$ in the training dataset leads, as it is seen, to a reduction of the uncertainty of the extrapolated value of $B_{\Lambda}$. Note also that successive extrapolates are consistent with the previous ones within the given uncertainties. One can conclude, therefore, that the extrapolated values of $B_\Lambda$ for the two states of both hypernuclei show a rather well convergence in terms of the maximal value of $N_{max}$, being this an indication that ANNs are a reliable method to extrapolate the results of hypernuclear NCSM calculations to large model spaces. We want to note now that whereas the prediction of our ANN for the $0^+$ state of $^4_\Lambda$He extrapolates, within the given uncertainty, rather well to the corresponding experimental result for large values of the model space size, the extrapolation for the two states of $^4_\Lambda$H and the $1^+$ state of $^4_\Lambda$He are off of the experiment by about 0.3 MeV. This discrepancy between our ANN prediction and the experimental results, however, should not be 
attributed to the performance of the ANN but to the Hamiltonian employed and the symmetries assumed in the NCSM calculations of Refs.\ \cite{Wirth18,Htun21,Gazda22}. We notice that the goal of the present work is mainly focused on  
discerning whether an ANN is a reliable scheme to extrapolate NCSM results at lager model spaces rather than on its accuracy on reproducing the experimental results. To such end, in Tab.\ \ref{tab:tab1}, we have compared the predictions of our ANN for the $\Lambda$ separation energies of the three hypernuclei considered here with the extrapolated results 
obtained in Refs.\ \cite{Wirth18,Htun21} for a regulator cutoff of $600$ MeV.  The interested reader is referred to Sec.\ III-C of Ref.\ \cite{Wirth18} and 
Sec.\ 2.2 of Ref.\ \cite{Htun21} for detailed descriptions of the extrapolation procedures employed in these two works. The experimental separation energies reported in Refs.\ \cite{Davis05,Gal16,Schulz16,Yamamoto15} is also shown. Our ANN predictions are shown for $N_{max}=100$ and the values of the HO spacing $\hbar\omega=10$ MeV for $^3_\Lambda$H and $\hbar\omega=30(32)$ MeV for the $0^+(1^+)$ states of $^4_\Lambda$H and $^4_\Lambda$He.  As it can be seen in the table our results are in excellent agreement with those of  Refs.\ \cite{Wirth18,Htun21}, being this is a clear indication that, as we already said, ANNs are a reliable method to extrapolate hypernuclear NCSM calculations to large model spaces. Finally, we would like to point out that our ANN does not explain the charge symmetry breaking (CSB) in the $A=4$ mirror hypernuclei. The reason being simply the fact that the NSCM calculations of Refs.\ \cite{Wirth18,Htun21,Gazda22}, used to train our ANN, do not include CSB effects and, consequently, our neural network cannot account for them.

\begin{table}[t]
\begin{center}
\begin{tabular}{c|ccc}
\hline
\hline
Hypernucleus &  ANN prediction  &  Extrapolated results of Refs.\ \cite{Wirth18,Htun21} & Exp.  \\
\hline
$^3_\Lambda$H (g.s.)& $0.16 \pm 0.02$  & $0.158$ \cite{Htun21} & $0.13 \pm 0.05$ \cite{Davis05,Gal16}\\ 
\hline
$^4_\Lambda$H $(0^+)$ & $2.49 \pm 0.05$ & $2.48 \pm 0.04$ \cite{Wirth18} & $2.157 \pm 0.077$ \cite{Schulz16} \\ 
$^4_\Lambda$H $(1^+)$ & $1.35 \pm 0.16$ & $1.40 \pm 0.28$ \cite{Wirth18} & $1.067 \pm 0.08$ \cite{Yamamoto15} \\ 
\hline
$^4_\Lambda$He $(0^+)$ & $2.43 \pm 0.06$ & $2.45 \pm 0.04$ \cite{Wirth18} & $2.39 \pm 0.05$ \cite{Davis05} \\ 
$^4_\Lambda$He $(1^+)$ & $1.33 \pm 0.11$ & $1.34 \pm 0.28$ \cite{Wirth18}& $0.984 \pm 0.05$ \cite{Yamamoto15} \\ 
\hline
\hline
\end{tabular}
\end{center}
\caption{Summary of the ANN prediction of the $\Lambda$ separation energies $B_\Lambda$ of the lightest hypernuclei. Results are shown for $N_{max}=100$ and the values of the HO spacing $\hbar\omega$=10 MeV for $^3_\Lambda$H and $\hbar\omega=30(32)$ MeV for the $0^+(1^+)$ state of $^4_\Lambda$H and $^4_\Lambda$He. The extrapolation of the hypernuclear NCSM calculations of Refs.\ \cite{Wirth18,Htun21} for a regulator cutoff of $600$ MeV are also shown for comparison. The experimental 
values, taken from Refs.\ \cite{Davis05,Gal16,Schulz16,Yamamoto15}, are shown in the last column. Units are given in MeV.}
\label{tab:tab1}
\end{table}

\section{Summary and Conclusions}
\label{sec:summ_conclu}

Using a feed-forward ANN we have extrapolated at large model spaces the results of the {\it ab-initio} hypernuclear No-Core Shell Model calculations of Refs.\ \cite{Wirth18,Htun21,Gazda22} for the $\Lambda$ separation energy of $^3_\Lambda$H, $^4_\Lambda$H and $^4_\Lambda$He. Due to the limited size of the input dataset, to avoid overfitting we have enlarged it by performing a cubic spline interpolation in the HO spacing $\hbar\omega$ at each given value of the model space size $N_{max}$ and, in addition, we have introduced a Gaussian noise in it during the training process of the network. We have found that an ANN with a single hidden layer made of eight neurons is enough to extrapolate correctly the $\Lambda$ separation energies of the three hypernuclei considered, in agreement with the {\it universal approximation theorem}, which states that any continuous function can be realized by a single-hidden-layer neural network.  
We have found that whereas the extrapolated results of the $\Lambda$ separation energy to large model spaces of size $N_{max}=100$ are in agreement with the experimental data in the case of $^3_\Lambda$H and the $0^+$ state of $^4_\Lambda$He, they are off of the experiment by about $0.3$ MeV for both the $0^+$ and $1^+$states of $^4_\Lambda$H and the $1^+$ state of $^4_\Lambda$He. This discrepancy between the ANN prediction and the experimental results should be attributed to the Hamiltonian employed and the symmetries assumed in the NCSM calculations of Refs.\ \cite{Wirth18,Htun21,Gazda22} and not to the ANN performance. The goal of the present work has been mainly focused on discerning whether an ANN is a reliable scheme to extrapolate NCSM results at lager model spaces rather than on its accuracy on reproducing the experimental results. We have found that our results are in excellent agreement with those obtained using the extrapolation schemes of Ref.\ \cite{Wirth18,Htun21} being this an indication that, an ANN is a reliable method to extrapolate the results of hypernuclear NCSM calculations to large model spaces.

To finish we would like to mention that it would be very interesting to compare our results with those obtained using other extrapolations schemes of the NCSM results such as that of infrared (IR) extrapolation \cite{Forssen18,Gazda22b} where the model space parameters $\hbar\omega$ and $N_{max}$ are translated into an IR length scale $L_{eff}$ and a ultraviolet (UV) cutoff scale $\Lambda_{UV}$. This comparison, however, is left for the near future. Another interesting point to be addressed in a future work is the analysis of the ANN performance on heavier hypernuclei. To the best of our knowledge NCSM calculations for heavier systems have been only performed  for the p-shell hypernuclei $^7_\Lambda$Li, $^9_\Lambda$Be and $^{13}_{\,\,\,\Lambda}$C \cite{Wirth14}, being the results less converged than those for the light hypernuclei. It would be, therefore, interesting to study how it is the convergence of the results for these heavier systems in the case of an ANN, and which are the pros and cons of this approach in comparison with the existing extrapolation methods.
 
\section*{Acknowledgments}

The author is very grateful to Daniel Gazda for providing him with the NCSM results used to train the neural network, to Avraham Gal and 
Dieterich Unkel for their useful comments, respectively, on the results obtained and the generalities of ML, and finally to Edoardo Lanza and Giuseppe Verde for their interesting discussions. This work has been supported by the European Union’s Horizon 2020 research and innovation programme under grant agreement No 824093.



\end{document}